\documentclass{jfm}
\usepackage{verbatim}
\usepackage{graphicx}
\usepackage{epstopdf, epsfig}
\usepackage{verbatim}
\usepackage{multirow}
\usepackage{bm}

\usepackage[colorlinks,
            linkcolor=blue,
            anchorcolor=blue,
            citecolor=blue,
            urlcolor=blue,
            ]{hyperref}

\shorttitle{Particle settling through a density transition layer}
\shortauthor{S. Wang, et al.}

\title{Bouncing behaviour of a particle settling through a density transition layer}

\author{Shuhong Wang\aff{1},
Prabal Kandel\aff{1},
Jian Deng\aff{1}
  \corresp{\email{zjudengjian@zju.edu.cn}},
  C.P. Caulfield\aff{2}
 \and Stuart B. Dalziel\aff{2}}

\affiliation{\aff{1}State Key Laboratory of Fluid Power and Mechatronic Systems, Department of Mechanics, Zhejiang University, Hangzhou 310027, People's Republic of China.
\aff{2}Department of Applied Mathematics $\&$ Theoretical Physics, University of Cambridge, Cambridge CB3 0WA, UK}

\begin{document}

\maketitle

\begin{abstract}
The present work focuses on a specific bouncing behaviour as a particle settling through a three-layer stratified fluid in the absence of neutral buoyant position, which was firstly discovered by \cite{abaid2004internal} in salinity-induced stratification. Both experiments and numerical simulations are carried out. In our experiments, illuminated by a laser sheet on the central plane of the particle, its bouncing behaviour is well captured. We find that the bouncing process starts after the wake detaches from the particle. The PIV results show that an upward jet is generated at the central axis behind the particle after the wake breaks. By conducting a force decomposition procedure, we quantify the enhanced drag caused by the buoyancy of the wake ($F_{sb}$) and the flow structure ($F_{sj}$). It is noted that $F_{sb}$ contributes primarily to the enhanced drag at the early stage, which becomes less dominant after the detachment of the wake. In contrast, $F_{sj}$ plays a pivotal role in reversing the particle's motion. We conjecture that the jet flow is a necessary condition for the occurrence of bouncing motion.
Then, we examine the minimal velocities (negative values when bounce occurs) of the particle by varying the lower Reynolds number $Re_l$, the Froude number $Fr$ and the upper Reynolds number $Re_u$ within the ranges $1 \leq Re_l\leq 125$, $115 \leq Re_u\leq 356$ and $2 \leq Fr\leq 7$. 
We find that the bouncing behaviour is primarily determined by $Re_l$.
In our experiments, the bouncing motion is found to occur below a critical lower Reynolds number around $Re^ \ast _{l}=30$. 
In the numerical simulations, the highest value for this critical number is $Re^ \ast _{l}=46.2$, limited in the currently studied parametric ranges.

\end{abstract}


\section{Introduction}

Density stratification caused by nonuniform distributions of temperature or salinity is ubiquitous in oceans and lakes, which alters the settling or rising process of submerging particles, and has marked effects on many environmental problems, such as the aggregation of marine snow, the formation of thin layers, the dispersion of spilling oil and the deposition of sediments \citep{prairie2017model,diercks2019vertical}.
Understanding the fluid dynamics of particle settling (or rising) in a stratified ambient fluid is of considerable significance for better predictions of those actions.

In homogeneous fluid, the hydrodynamic force acting on a particle accelerating under gravity can be decomposed into buoyancy force, steady-state drag, added mass, and history (Basset) force. The steady-state drag increases with velocity and finally balances the reduced gravity, such that the particle reaches a steady state. In stratified fluid, the particle experiences an additional drag force due to the stratification, noted as `stratification drag', yielding a significant decay in settling velocity \citep{srdic1999gravitational,abaid2004internal,camassa2009prolonged,2009Enhanced,camassa2010first,doostmohammadi2014numerical,verso2019transient,mandel2020retention,magnaudet2020particles}. 

There have been different explanations for the origin of stratification drag. A widely accepted one is that it comes from the buoyancy of associated upper lighter fluid, as the settling particle distorts the isopycnals and drags some upper fluid to a lower position.
\citet{srdic1999gravitational} conducted experiments of particles settling in a three-layer stratified fluid, with two homogeneous layers and one density transition layer (interface) in between, at Reynolds numbers $1.5\leq Re \leq 15$ and Froude numbers $3\leq Fr \leq 10$. They found that the total drag enhancement can be estimated by the total buoyancy of the upper fluid dragged below the upper bound of the interface, before the maximum drag is reached. The contribution of internal waves is negligible before the wake breaks as they are generated after the rupture of the wake.
With the same type of stratification,
\citet{verso2019transient} studied experimentally the motion of four different particles, in a wider parameter range ($2\leq Re \leq 106$ and $0.5 \leq Fr \leq 28$). A time-dependent stratification force model was developed for those particles, based on the assumption that the stratification force is entirely contributed by the buoyancy of an effective wake, and the wake volume is constant within the interface and decreases exponentially till a new terminal velocity is reached.

In a linearly stratified fluid, a combined experimental and numerical investigation of settling particles at small Reynolds numbers ($Re \sim O(1)$) was presented by \citet{2009Enhanced}. They found that the buoyancy of a shell of fluid around the particle, instead of the entire distorted region, is responsible for the drag enhancement. Furthermore, they suggested that the total drag increment can be scaled by a dimensionless parameter, the Richardson number, characterising the relative importance of buoyancy and viscous shear force. 
A similar mechanism was also found for the stratification drag of a rising grid of bars \citep{higginson2003drag}, at much higher Reynolds numbers ($Re \sim O(10^3)$), using the drift volume as an approximation of the volume of dragged fluid. 

Although the associated buoyancy accounts for the stratification drag in many cases, \citet{zhang2019core} pointed out that it is the specific structure of the vorticity field induced by the buoyancy effects that contributes mainly to the stratification drag, while the buoyancy itself plays a secondary role. 
Moreover, \citet{2000Flow} noted that the drag enhancement is rather small until a vertical upward jet is generated at the rear of the particle at relatively strong stratification ($Fr \lesssim 20$), evidencing the drag contribution from the flow structure. 

The flow structure manifested by the vertical motion of a particle in a stratified fluid is notably different from that in a homogeneous fluid. In a stratified fluid, the baroclinic torque ($\bigtriangledown \rho \times \bigtriangledown p $) results in vorticity generation whenever there is a misalignment between density and pressure gradient \citep{magnaudet2020particles}.
For low Reynolds numbers ($Re\ll 1$), top-down symmetrical toroidal eddies are induced by a vertical, downward point force (Stokeslet) under linear stratification, which is similar to the eddy formed under the restriction of two horizontal walls, indicating the suppression of vertical flow by stratification \citep{ardekani2010stratlets}.
For higher Reynolds numbers ($0.05\leq Re \leq100$), toroidal eddies still exist but lose the top-down symmetry, and a vertical upward jet is generated simultaneously at the centre line downstream of the particle \citep{zhang2019core}.  This jet was experimentally observed over a wide range of Froude numbers, $0.2 \leq Fr \leq 70$, and Reynolds numbers, $30 \leq Re \leq 4000$, with its shape and strength varying with $Re$ and $Fr$ \citep{hanazaki2009jets}. 

The present work focuses on the transient bouncing behaviour as a particle passes through a density transition layer with a large density gradient, while the particle density is always higher than the fluid such that no neutral position exists. This phenomenon was first observed by \citet{abaid2004internal} in their experiments with a strong salt stratification. As we know that in continuous concentration-induced stratified fluid, a droplet could bounce and oscillate under the Maragoni effects induced by nonuniform surface tension \citep{blanchette2012drops,li2019bouncing}. However, this mechanism could not be applied to a solid particle, as surface tension does not exist at a solid-liquid surface. \citet{abaid2004internal} observed that before the particle bounces up, the entrained plume detaches and ascends to the upper layer, which indicates that the particle is possibly lifted by the plume.
However, the detached plume moving opposite to the particle is not uniquely associated with the bouncing behaviour. A descending wake of a droplet rising through a density interface could not trigger a reverse motion \citep{mandel2020retention}. Moreover, wake rupture was also observed experimentally for a descending solid particle \citep{srdic1999gravitational}.
Therefore it is of interest to further explore the mechanisms accounting for the bouncing of a solid particle.

As the bouncing behaviour occurs at certain combinations of parameters, a parametric study is necessary, before we further investigate the bouncing mechanisms. Previous work revealed that the bouncing behaviour is affected by a variety of parameters.
\citet{camassa2022critical} found that the critical particle density $\rho_p$ for the occurrence of bouncing can be expressed by a linear combination of upper and lower layer fluid densities. \citet{doostmohammadi2014reorientation} found that at a relatively higher density ratio of $(\rho_l-\rho_u)/\rho_u$, an ellipsoid could bounce for a short time as it passes through a density interface. Additionally, \citet{blanchette2012drops} found that high ratio of $(\rho_p-\rho_u)/(\rho_p-\rho_l)$ could lead to a temporary reverse motion of a drop as it settles through a density transition layer with identical surface tension. 
These studies indicate that the bouncing behaviour is strongly dependent on the density or density ratio of the system. Moreover, the Froude number is a key parameter affecting the oscillation of a particle or droplet near their neutral buoyant position in a linear stratified fluid \citep{bayareh2013rising,doostmohammadi2014numerical}. In the experiments carried out by \citet{abaid2004internal}, the transient levitation of the particles was discovered by adjusting the lower terminal velocities, which means that the lower Reynolds number should be taken into account.
As we know, in three-layer stratification, both the Froude number and the Reynolds number are dependent on the density difference. Whether the controlling parameter of the bouncing motion is the density difference, or other relevant parameters, is necessary to be addressed.

The paper is organised as follows:
In section \ref{sec:exp_approach}, we introduce the experimental method, including experimental setup and measurement. The numerical method and validation are presented in section \ref{sec:numerical}.
We discuss our results in section \ref{sec:results}. We present the settling process of a typical bouncing particle, and analyse the forces acting on it, to understand the mechanisms of bouncing behaviour. Then we examine the effects of $Re_l$, $Fr$ and $Re_u$ on the minimal velocity of the particle, to identify the key controlling parameter. Conclusions are drawn in section \ref{sec:conclusions}.

\begin{figure}
  \centerline{\includegraphics[width=0.8\textwidth]{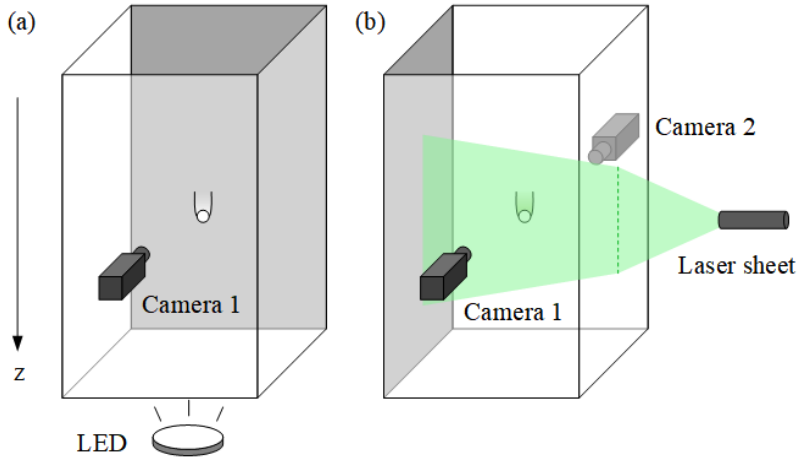}}
  \caption{Experimental setups. (a) Setup 1 for recording particles trajectories; (b) Setup 2 for wake visualisation and flow fields measurement.}
\label{fig:setup}
\end{figure}

\section{Experimental approach}\label{sec:exp_approach}
\subsection{Experimental setup}
The experiments were performed in a glass tank,  with a total depth of 60 cm to ensure that terminal velocity was achieved. The tank has a large base area (30 cm $\times$ 30 cm) to avoid interaction between the particle and the tank walls. To prepare the working fluids, salt is dissolved in fresh water with different concentration ratios in several separate buckets. They are then kept at room temperature for at least 24 hours before filling the tank, to eliminate resolved gas and get uniformly stable concentration distribution. 

The tank is first half-filled by heavy fluid. Then light fluid is pumped slowly and horizontally to the top of the heavy fluid by a micropump, at a low flow rate of 100 ml/min, to minimise mixing of the two fluids. This filling method yields an error-function-type density profile. The tank is let stand for at least half an hour after the filling, and before the experiments, to diminish the disturbances caused by the pumping. The standing time varies with cases to get different thickness of the transition layer, i.e., thicker transition layer needs longer time for the interdiffusion of two fluids.

Nylon particles are used for the experiments, with diameter $D=10$ mm and densities ranging from 1121-1126 kg/cm$^3$. The particles are released by a fixed clamp, at the centre of the tank cross-section (15 cm from the side walls), 2 cm below the free surface, and approximately 25 cm above the density transition layer. The particles are retained in another tank of stratified fluid at the same room temperature before release. Particle density measurements are conducted just before the experiments.
The time interval between each release is 10 minutes, ensuring that the fluid is relaxed to its quiescent state. The settling processes of particles are captured by a high speed camera at a frame rate of 100 fps.

Two different experimental setups are used, as shown in figure \ref{fig:setup}, to meet the requirements respectively for recording the trajectories of particles, and measuring the surrounding flow fields. For trajectory recording, we simply position a single camera at one side of the tank, and illuminate the tank via a panel of light emitting diodes (LEDs) from the bottom (figure \ref{fig:setup}(a)). The opposite sidewall of the tank is 
brushed black paint to avoid light reflection. The size of camera captured window is about 20 $\times$ 20 cm, yielding a resolution of $0.2$ mm/pixel. 

In the second experimental setup, the upper fluid is dyed using Rhodamine B for wake visualisation, and seeding particles are added to both the upper and lower layer fluids for Particle Image Velocimetry (PIV) measurement. To simultaneously capture the visualised wake structure and measure the velocity fields, two cameras are placed at opposite sides of the tank and carefully adjusted to get their optical axis parallel (figure \ref{fig:setup}(b)). The centre plane of the tank which is perpendicular to the camera optical axis is illuminated by a laser sheet (thickness $\sim$ 2 mm) sideways. The two cameras are synchronised to obtain simultaneous image pairs. One camera is equipped with a long-pass filter to capture the dyed wake, and the other one with short-pass filter to get the images of seeding particles. 
To capture the flow structure of the whole settling process, the particle should stay within the illuminated laser sheet plane. The densities of the particle and fluid should be carefully chosen to avoid out-of-plane motion, and the particle should be released very carefully to minimise disturbances. 
It is worth mentioning that the laser sheet hits the particle surface and heats the vicinity fluid. This effect is negligible in the upper layer as the particle settles fast but prolongs the suspending time after it enters the lower layer. For flow visualisation and PIV measurements, where we focus on the flow structure, the laser sheet is still an effective illumination approach.

\begin{figure}
  \centerline{\includegraphics[width=1\textwidth]{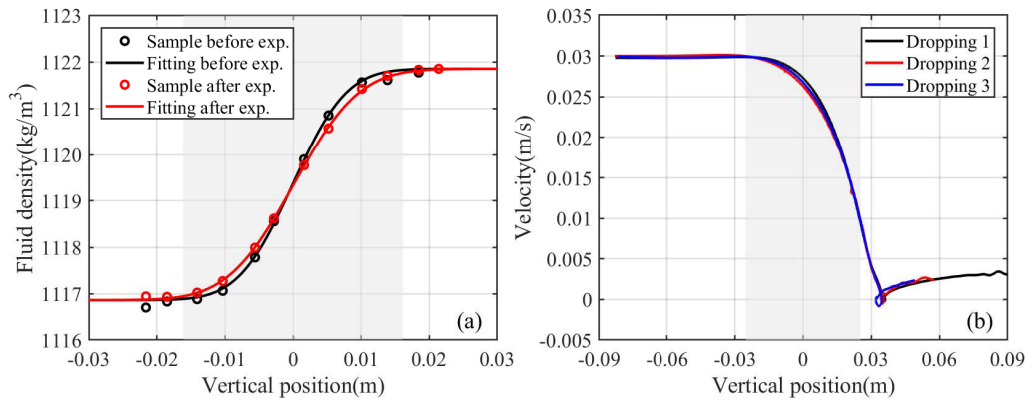}}
  \caption{(a) Density distribution before and after one experiment. The grey area represents the density interface, which covers $98\%$ density variation between the upper and lower layers. (b) Velocity profiles of three repeated droppings of a same particle.}
\label{fig:density_repeat}
\end{figure}

\subsection{Experimental measurements}
The fluid density is measured by a densitometer (Anton-Parr DMA 4500) with measuring accuracy of 0.01 kg/m$^3$. To measure the particle density with the same accuracy, the particle is suspended in a tank ($10\times 10 \times50$ cm) filled with the fluid of an approximately linear density profile. The sampled fluid at the same level of the suspended particle is then measured, of which the density is taken as the particle density. Particle diameters are measured by a micrometer with accuracy of $0.001$ mm. 

To measure the density distribution of the transition layer, 12 thin needles are inserted horizontally into the tank from punched holes at the sidewall, with 4 mm vertical intervals, and connected to 12 syringes for taking samples.
Each sample takes 2.5 ml fluid. For each measurement, 30 ml fluid is taken in total, which leads to a variation in height of less than 0.4 mm, for the working tank of cross-section 30 cm $\times$ 30 cm. 
Thus the modification to the density distribution by the measurement can be ignored. 
Density distribution measurements are performed twice for each test, before and after dropping the particles, respectively. 

The measured density of the 12 samples are fitted to an error-function-shaped function, satisfying the following expression:
\begin{equation}
  \rho = \frac{\rho_u+\rho_l}{2} + \frac{\rho_u-\rho_l}{2} \mbox{erf}(\alpha(h-h_{ref})),
  \label{rho}
\end{equation}
where $\rho_u$ and $\rho_l$ are respectively the upper and lower layer fluid densities, $h_{ref}$ is the reference height, corresponding to the centre of the density transition layer, $\alpha$ is a scaling factor, determining the thickness of the density transition layer, and erf($x$) is the error function, which is written as
\begin{equation}
\mbox{erf}(x)=\frac{2}{\pi} \int_{0}^{x}e^{-t^2}dt.
  \label{erf}
\end{equation}
The measured density profiles for one of the experiments are presented in figure \ref{fig:density_repeat}(a). We can see that the density profile becomes smoother after the experiment.
The average of these two measurements is then used as the final density profile.

It is important to monitor and control the temperature of working fluid.
With a salinity of 16$\%$ and temperature of $18 ^{\circ}$C (according to the present working fluid), $1^{\circ}$C of temperature variation can lead to 0.41 g/cm$^3$ density variation of the fluid. The natural room temperature variation from daytime to nighttime can be up to $10^{\circ}$C, which may significantly alter the particle behaviour. To minimise the temperature variation of the working fluid, the experiments are conducted in an enclosed room with the room temperature controlled by an air conditioner. 
The real-time temperature is monitored during each test, and the temperature variation for all tests is maintained at less than $1^{\circ}C$. 
The density measurement for each particle is conducted just before the dropping to minimise the influence of temperature.

The viscosity of fluid is calculated by the following empirical formula with accuracy of $\pm 1.5\%$ (see equation 22 in the reference by \citet{sharqawy2010thermophysical}):
\begin{equation}
  \left\{
    \begin{array}{l}
      \mu = \mu _w(1+a_1S_a+a_2S_a^2),\\
      \mu_w=4.2844\times10^{-5}+0.157(T_e+64.993)^2-91.296)^{-1}, \\
      a_1=1.541+1.998\times10^{-2}T_e-9.52\times10^{-5}T_e^2  \\
      a_2=7.974-7.561\times10^{-2}T_e+4.724\times10^{-4}T_e^2,
  \end{array} \right.
  \label{nu}
\end{equation}
where $S_a$ and $T_e$ are salinity and temperature respectively. 

Instantaneous particle displacements are obtained by fitting the discrete time-dependent position points with a cubic smoothing spline, following the methods of \citet{truscott2012unsteady} and \citet{epps2010impulse}. The fitting error tolerance is E=1e-7 for all experimental position data, which provides accurate fitting data and gives smooth fitting derivatives.
Then, the velocity and acceleration of the particles can be calculated based on the time histories of the fitted particle displacements.

Since the passing of particles will introduce disturbances to the density transition layer and expedite the diffusion, no more than five particles are dropped in each fluid tank.  
The repeatability validation is conducted separately before the experiment. As presented in figure \ref{fig:density_repeat}(b), the settling dynamics is almost unchanged for the repeated releases. 

\section{Numerical method}\label{sec:numerical}

We solve the time-dependent incompressible Navier-Stokes equations with finite volume method. The continuity and the momentum equations are expressed as
\begin{equation}
\nabla \cdot \bm{u} =0,
  \label{eqn:ns1}
\end{equation}
\begin{equation}
\rho(\frac{\partial \bm{u}}{\partial t}+ \bm{u} \cdot \nabla \bm{u})= -\nabla p + \mu \nabla ^2 \bm{u} +\rho \bm{g}.
  \label{eqn:ns2}
\end{equation}
The Boussinesq approximation is applied to account for the stratification effect, where the density variation enters the momentum equation only through the buoyancy term. Division by the reference density in (\ref{eqn:ns2}) yields
\begin{equation}
    \frac{\partial \bm{u}}{\partial t}+ \bm{u} \cdot \nabla \bm{u}= -\frac{1}{\rho_{ref}} \nabla p + \frac{\rho}{\rho_{ref}}\bm{g} + \nu \nabla ^2 \bm{u},
    \label{eqn:ns3}
\end{equation}
with kinematic viscosity $\nu=\mu/\rho_{ref}$. The transport equation for density is given as
\begin{equation}
\frac{\partial \rho}{\partial t}+ \bm{u} \cdot \nabla \rho= \kappa \nabla ^2 \rho,
  \label{eqn:ns4}
\end{equation}
where $\kappa$ is the scalar diffusivity defined as $\kappa=\nu/Pr$. In our simulations, we choose Prandtl number $Pr=700$, that corresponds to the salinity-induced stratification in water. We note that for stratified flows with $Pr>1$, the scales are smaller for density than the velocity at a given $Re$. To resolve the dynamic scales in such a stratified flow, very fine spatial resolution is required \citep{orr2015numerical}. In order to estimate the momentum and density boundary layer thickness ($l_{m}$ and $l_{d}$ respectively), we adopt the following relations \citep{schlichting2003boundary}:
\begin{equation}
 l_{m} \sim O\left(\frac{d}{\sqrt{R e}}\right)
 \label{eq:lm}
\end{equation}
and
\begin{equation}
l_{d} \sim O\left(\frac{d}{\sqrt{{\it{Re Pr}}}}\right).
\label{eq:ld}
\end{equation}
The moderate Reynolds numbers considered in the present work ($Re\leq356$) allow the application of axisymmetry assumption. A three-dimensional numerical simulation of a particle (sphere) settling in linearly stratified fluid shows that the flow remains axisymmetric at Reynolds number up to 356 without vortex shedding \citep{doostmohammadi2014numerical}. It has also been reported that nearly axisymmetric structure is retained even at $Re\sim 800$ \citep{2000Flow}. 
After testing different spatial resolutions, we find that the mesh with approximately 348,000 cells provides satisfactory accuracy. The first cell layer around the particle surface is with the height of $0.0014D$, which grows exponentially at a slow rate normal to the sphere surface. The grid size grows from the surface to a maximum cell length of $0.1D$ at the radial distance $10D$ from the particle centre, and then keeps constant to the outer boundaries.
According to (\ref{eq:lm}) and (\ref{eq:ld}), $l_{m} \sim 0.053D$ and $l_{d} \sim 0.002D$ for $Re=356$ and $Pr=700$, respectively. Note that the choice of $Re$ here corresponds to the highest $Re$ in our numerical simulations. Though the mesh resolution we adopt satisfies $l_{d}$ around the vicinity of the sphere surface, we note that the resolution in the far wake region may not resolve the density gradient well. It would be noteworthy to mention that a numerical study by \citet{2000Flow} suggested that in a moderate Reynolds number regime, the actual thickness of the density boundary layer can be larger than the predicted value of (\ref{eq:ld}). This was also verified by \citet{doostmohammadi2014numerical} in their simulations of a settling sphere in a linearly stratified fluid with $Pr=700$. The authors carried out tests with minimum cell length up to 6 times larger than those predicted from (\ref{eq:ld}), and observed little deviation in their results. While high computational cost is a limiting factor for global density boundary layer resolution in the domain, a qualitative examination still shows that our numerical simulations capture the structures evolving in the stratified wake to a good extent.

We solve the governing equations of the stratified flow using finite-volume based code \citep{weller1998tensorial}. The space discretisations are second-order upwind for the convection terms and
central differences for the Laplacian terms, respectively. The time discretisation
is second-order implicit Euler. The pressure-velocity coupling is obtained using the PISO (Pressure-Implicit with Splitting of Operators) scheme. For the velocity boundary conditions, we set that on the sphere surface to be moving-wall with no flux normal to the wall. A zero-gradient condition is imposed on all other velocity boundaries. The pressure is fixed at the top boundary. A fixed-flux condition is adopted for the pressure at all other boundaries, such that the velocity boundary condition can specify the flux on those boundaries. The density is fixed at the top and bottom boundaries, with constant values ($\rho_{u}$ and $\rho_{l}$ respectively), and a zero-gradient condition for density is imposed on other boundaries. 

\begin{figure}
  \centerline{\includegraphics[width=1\textwidth]{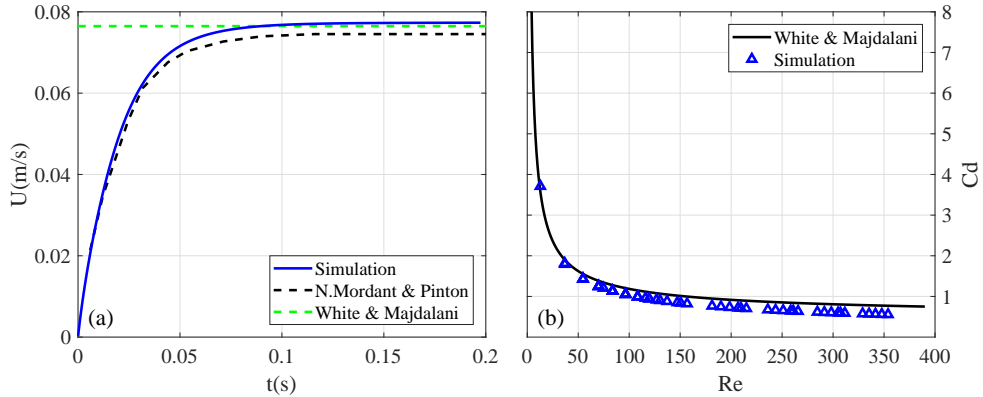}}
  \caption{(a) Velocity profiles of a particle settling in a quiescent homogeneous fluid at $Re=41$. (b) Drag coefficients for particles settling in homogeneous fluid.}
\label{fig:vali_numerical}
\end{figure}

The numerical methods are tested against experimental results of a particle settling in homogeneous and stratified fluids. The velocity profile of a particle settling in a quiescent homogeneous fluid is compared with the experimental results of \cite{mordant2000velocity}, as shown in figure \ref{fig:vali_numerical}(a). The velocity matches the experiment well at the beginning, but is a little bit higher after the particle reaches a steady state. The terminal velocity is dependent on the steady-state drag, which is defined as 
\begin{equation}
F_d=-\frac{1}{2} C_d \rho_f U|U| S_p,
  \label{eqn:Fd}
\end{equation}
where $C_d$ is the drag coefficient and $S_p$ is the cross-sectional area of the particle. $C_d$ can be calculated by an empirical formula 
\begin{equation}
C_d=\frac{24}{R_e}+\frac{6}{1+\sqrt{Re}}+0.4,
  \label{eqn:Cd_white}
\end{equation}
with error less than $10\%$ for $0< Re< 2\times 10^5$ \citep{white2006viscous}. The simulated drag coefficients are presented in figure \ref{fig:vali_numerical}(b), which qualitatively agree with that predicted by (\ref{eqn:Cd_white}). 
Further, We test the numerical method with a particle settling in stratified fluid. The transient flow structures are shown in figure \ref{fig:sim_process}, comparing to the experiment presented in figure \ref{fig:piv}. The bouncing behaviour is well captured, as shown in figure \ref{fig:valid_vy}. The results show that the simulation results agree well with our experiments. Those numerical results will be discussed later in detail.

\section{Results and discussion}\label{sec:results}

\begin{table}
  \begin{center}
\def~{\hphantom{0}}
  \begin{tabular}{llll}
      Parameter           & Symbol     &  Definition        & Range of values \\ 
      Particle density    & $\rho_p$   & $-$          & $1121.58 \sim 1125.26$ kg/m$^3$  \\ 
      Particle diameter   & $D$        & $-$          & $10.055 \sim 10.128$ mm  \\ 
      Particle velocity   & $U$        & $-$          & $-0.559 \sim 4.908$ cm/s  \\ 
      Jet velocity & $u_j$        & $-$          & $-$  \\ 
      Upper fluid density & $\rho_u$   & $-$          & $1113.92 \sim 1118.63$ kg/m$^3$  \\ 
      Lower fluid density & $\rho_l$   & $-$          & $1121.45 \sim 1123.66$ kg/m$^3$  \\ 
      Interface thickness & $L$        & $-$          & $2.52 \sim 13.06 $ cm \\ 
      Upper Reynolds number & $Re_u$   & ${\rho_u U_uD}/{\mu_u}$          & $115 \sim 356 $ \\
      Lower Reynolds number & $Re_l$   & ${\rho_l U_lD}/{\mu_l}$          & $1 \sim 125$  \\ 
      Froude number         & $Fr$   & $U_u/ND$          & $2 \sim 7 $ \\
      Brunt-V{\"a}is{\"a}l{\"a} frequency & $N$     & $\sqrt{(g(\rho_l-\rho_u)/L\rho_{ref})}$          & $0.62 \sim 1.88 $ \\
      Reference density     & $\rho_{ref}$   & $(\rho_u+\rho_l)/2$          & $1118.08 \sim 1120.09 $ kg/m$^3$  \\
      Prandtl number        & $Pr$     & $\nu/\kappa$  & $\sim700$  \\
      density ratio         &$\Delta \rho_l$ &$(\rho_p-\rho_l)/\rho_l$ &  $(0.03 \sim 2.54) \times10^{-3}$\\
  \end{tabular}
  \caption{Definition and ranges of parameters covered in the present work.}
  \label{tab:para}
  \end{center}
\end{table}

\subsection{Controlling parameters}
A particle settling in a stratified fluid is mainly characterised by three non-dimensional parameters, the upper layer Reynolds number $Re_u=\rho_u U_u D/\mu_u$, the lower layer Reynolds number $Re_l=\rho_l U_l D/\mu_l$, and the Froude number $Fr=U_u /ND$, where $U$ is the terminal settling velocity of the particle in a homogeneous upper or lower fluid, $\mu$ is the dynamic viscosity of the fluid, $D$ is the particle diameter, and the subscripts $u$ and $l$ represent respectively the upper and lower layer fluids. The Brunt-V{\"a}is{\"a}l{\"a} frequency $N$ is calculated as 
\begin{equation}
N=\sqrt{\frac{2g}{\rho_u +\rho_l}\frac{\rho_l -\rho_u}{L}},
  \label{eqn:BruntFrequency}
\end{equation}
where $L$ is the thickness of the density transition layer, covering $98\%$ of the density variation (figure \ref{fig:density_repeat}(a)).
All relevant parameters are listed in table \ref{tab:para}.

\begin{figure}
  \centerline{\includegraphics[width=1\textwidth]{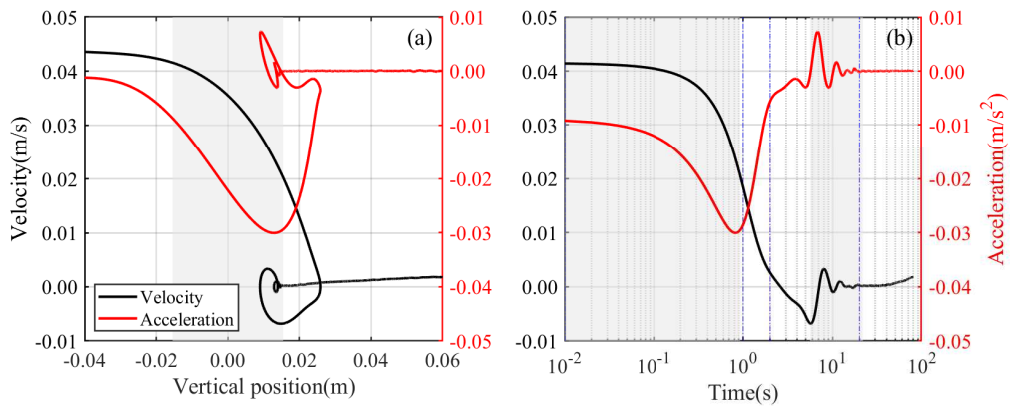}}
  \caption{Settling velocity and acceleration of the particle with $Re_u=347$, $Re_l=20$ and $Fr=2.5$ as the functions of  (a) the vertical position and (b) settling time. The left and right vertical axis correspond respectively to the settling velocity and acceleration. In (a), the vertical position ($z=0$) refers to the middle plane of transition layer, and the time $t=0$ refers to the instant when the particle's centre reaches the upper boundary of the transition layer. In (a), the shaded region represents the transition layer, and in (b), the two shaded regions denote when the particle's centre is within the transition layer. Note the right-side shaded region in (b) denotes when the particle re-enters the transition layer after bouncing. In (b), four stages are divided by blues dashed lines.}
\label{fig:va0325}
\end{figure}

\begin{figure}
  \centerline{\includegraphics[width=0.9\textwidth]{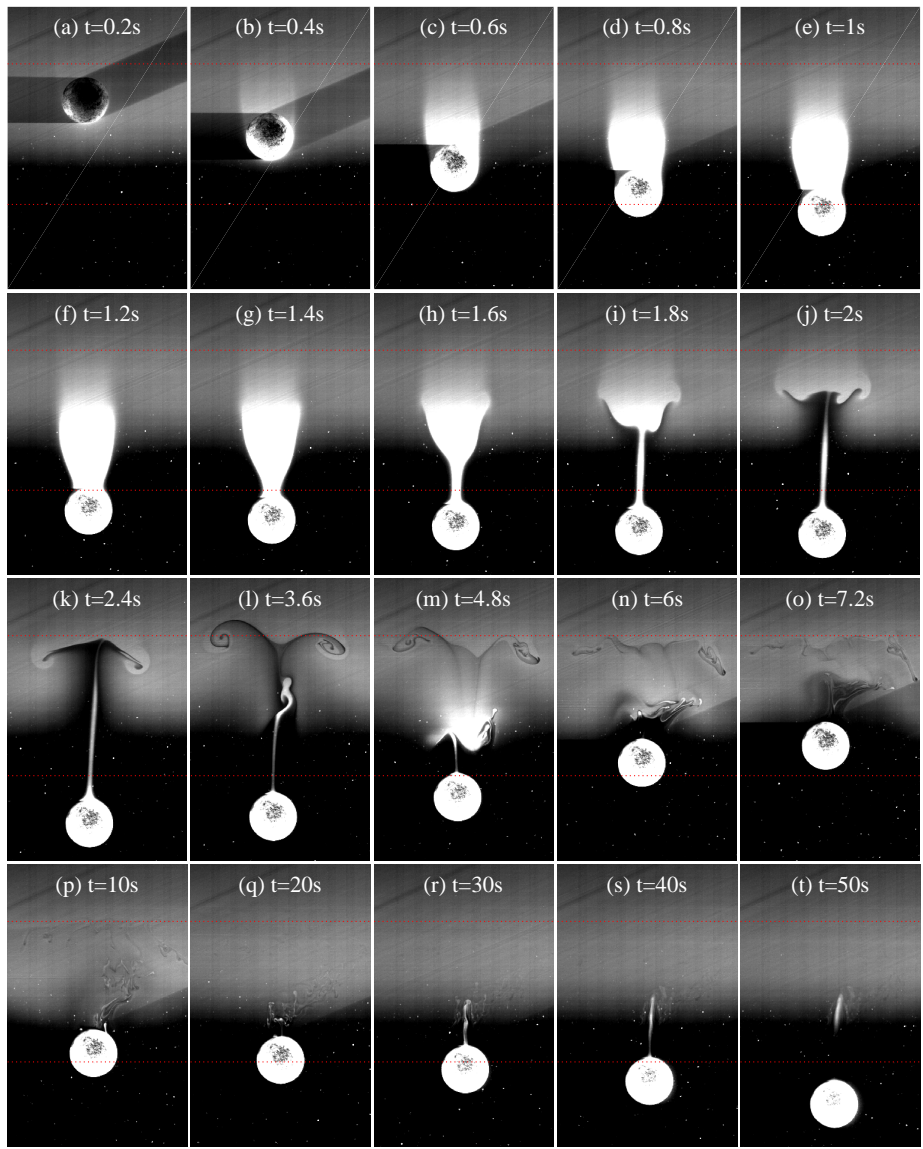}}
  \caption{A sequence of images showing the bouncing process of a particle settling through a density transition layer, corresponding to figure \ref{fig:va0325}. The two red dashed lines in each image represent the bounds of the interface. The non-dimensional parameters are $Re_u=347$, $Re_l=20$ and $Fr=2.5$.}
\label{fig:bounce}
\end{figure}
\begin{figure}
  \centerline{\includegraphics[width=0.8\textwidth]{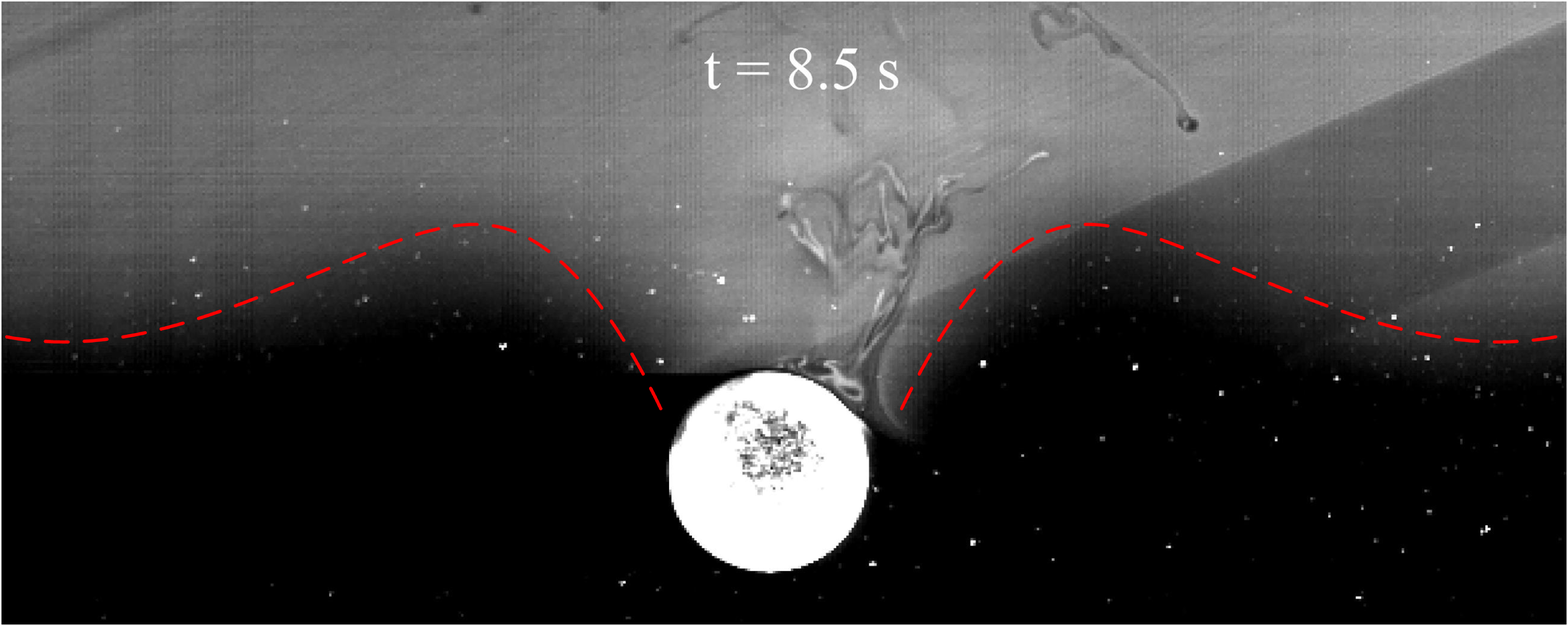}}
  \caption{Full image of internal wave at $t=8.5$s between figure \ref{fig:bounce}(o) and \ref{fig:bounce}(p).}
\label{fig:wave}
\end{figure}

\subsection{Bouncing process}\label{sec:bounce_process}
First, we choose a typical set of parameters, to demonstrate the bouncing process, or levitation \citep{abaid2004internal}, as a particle settles through a density transition layer. The velocity and the acceleration, as well as the visualised flow, of a classical bouncing particle are presented.
Here, the non-dimensional parameters are $Re_u=347$, $Re_l=20$ and $Fr=2.5$. 
At these Reynolds numbers, the flow is nearly axisymmetric thus the particle can stay in the laser-sheet-illuminated plane and be captured during the whole settling process. 

The velocity and acceleration of the bouncing particle are presented in figure \ref{fig:va0325}. The deceleration begins before the particle enters the density interface (figure \ref{fig:va0325}(a)), as the existence of the interface limits the vertical excursion of fluid \citep{ardekani2010stratlets}. As the particle enters the interface, the velocity decreases more rapidly.
The maximum deceleration is reached at approximately the lower bound of the interface. 
The particle reaches a zero velocity in the lower layer and bounces up, reentering the interface. 
The bouncing is like a beginning of a damped oscillation, which is more obvious in figure \ref{fig:va0325}(b). Note that the triggered oscillation is not a common feature for a bouncing particle. It occurs only at strong bouncing where the particle could reenter the interface. 
Finally, the particle accelerates again and slowly approaches its new terminal velocity. 

A series of images corresponding to figure \ref{fig:va0325}, showing the development of the wake over the whole settling process, are presented in figure \ref{fig:bounce}.
We divide the settling process into four stages, and separate them by vertical dashed lines in figure \ref{fig:va0325}(b).
(1) Wake attachment (figure \ref{fig:bounce}(a)-(e)). The particle enters the interface and drags a large amount of upper fluid at its rear. The velocity decreases rapidly and the deceleration reaches its maximum at the end of this stage (figure \ref{fig:va0325}(b)).  
(2) Wake detachment (figure \ref{fig:bounce}(f)-(j)). At this stage, most of the attached upper fluid (wake) detaches from the particle and returns to its neutral position, although it could hardly return to the upper layer as its density has been increased by the mixing. 
At the centre axis above the particle, a long thin column of lighter fluid keeps attached.
This column does not break but elongates and gets thinner until it is too thin to be captured. The velocity continues to decrease and the acceleration becomes smaller, compared to the first stage (figure \ref{fig:va0325}(b)). By the end of this stage, the particle loses over $90\%$ of its entering velocity $U_u$.
(3) Transient bouncing (figure \ref{fig:bounce}(k)-(q)). The particle reaches a zero velocity ($t=2.61s$) and bounces up. 
Triggered by the rupture of wake, strong internal wave is generated at the interface (figure \ref{fig:wave}), which causes oscillation of the particle. At this stage, almost all of the lighter fluid has detached from the particle, which indicates that the bouncing of particle is more likely induced by the flow, or specific flow structure, not the buoyancy of wake. More details of the bouncing mechanism will be discussed in section \ref{sec:force_analysis}. 
(4) Final sedimentation (figure \ref{fig:bounce}(r)-(t)). After the bouncing and oscillation, the particle slowly settles to the bottom of tank. Note the increased time interval between images. The particle settles extremely slowly after passing through the density interface. 

These four stages are typical stages that a bouncing particle would experience. Note that the bouncing process begins at the third stage, after the wake detaches from the particle. This is in agreement with the phenomena observed by \citet{abaid2004internal} that the particle changes its moving direction after the ascending of the `plume', where the `plume' is actually the detached wake in our experiment. For a monotonously settling particle, the first two stages are same, followed by a final sedimentation without bouncing up.

\subsection{Flow structure}\label{sec:flows}
\begin{figure}
  \centerline{\includegraphics[width=0.9\textwidth]{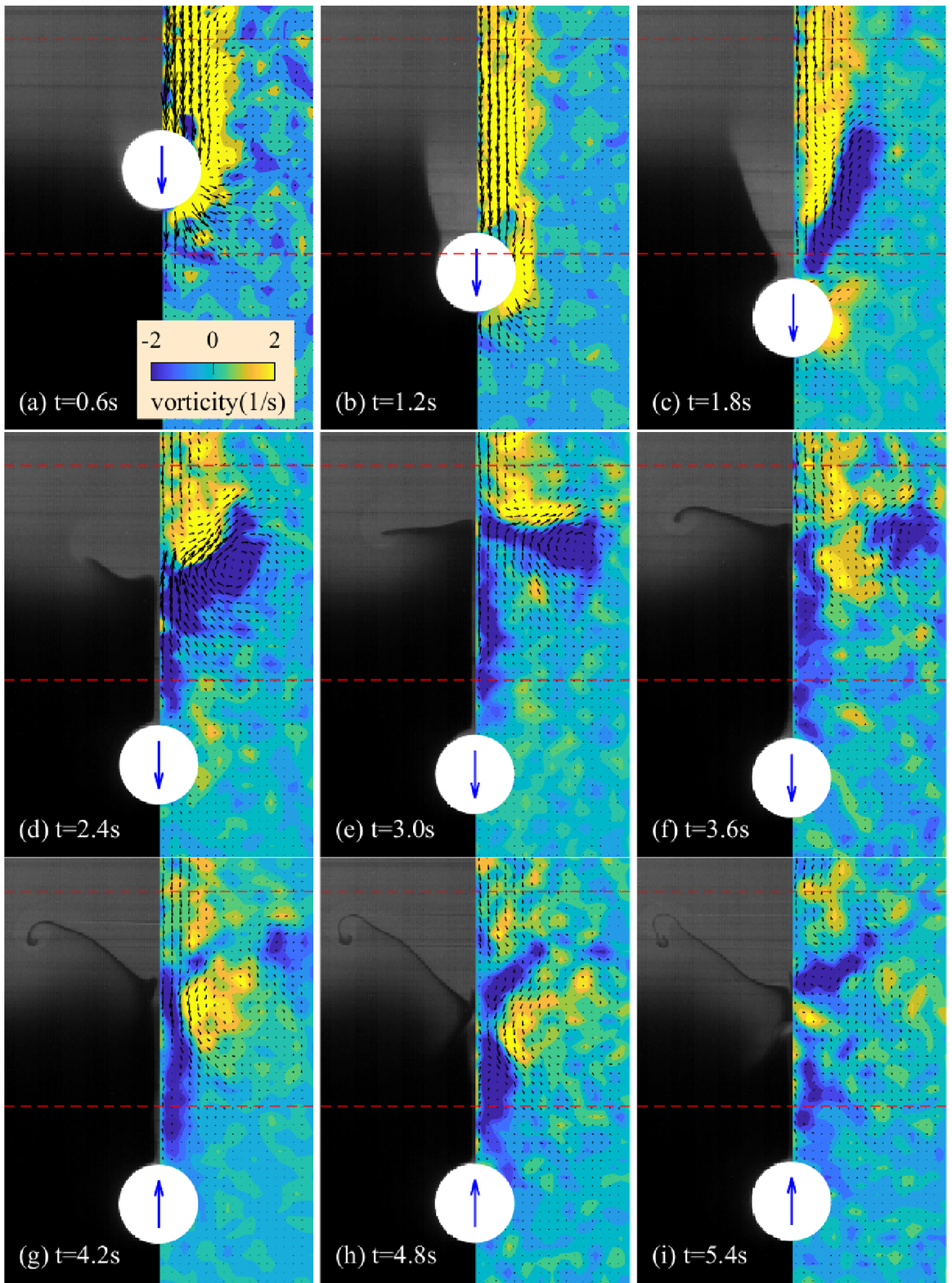}}
  \caption{Illuminated wake (left) and PIV fields (right) of a particle settling through a density transition layer at $Re_u=198$, $Re_l=20$ and $Fr=2.3$. The horizontal red dashed lines mark the bounds of the interface. The arrows at the particle centre point to the moving direction of the particle.}
\label{fig:piv}
\end{figure}

\begin{figure}
  \centerline{\includegraphics[width=0.9\textwidth]{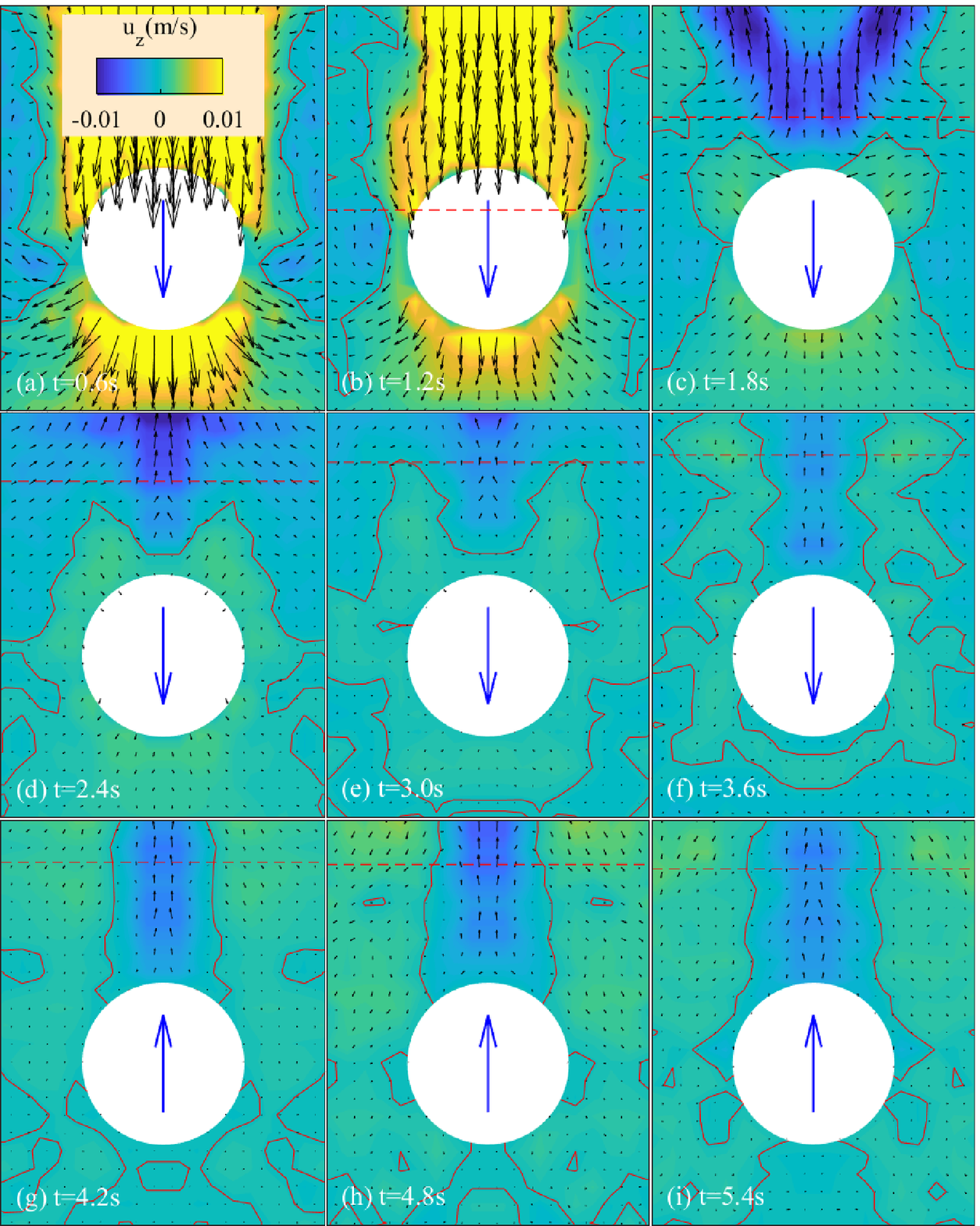}}
  \caption{A close view corresponding to the vicinity around the particle of figure \ref{fig:piv}. The red solid lines are isolines of $u_z=0$. A upward jet is generated at the centre axis behind the particle.}
\label{fig:piv_particle}
\end{figure}


To measure the transient flow structure around the particle, PIV and wake visualisation images are captured by two cameras simultaneously, using the experimental setup 2 illustrated in figure \ref{fig:setup}.
In figure \ref{fig:piv}, we present the results of a bouncing particle with non-dimensional parameters $Re_u=198$, $Re_l=20$ and $Fr=2.3$. 
Initially, the flow around the particle is similar to that in a homogeneous fluid. 
After entering the interface, accompanying the rupture of the wake, a vortex with direction opposite to that in the homogeneous fluid, is formed at the rear of the particle. It quickly grows and detaches from the particle, becomes a vortex ring remaining at the interface (figure \ref{fig:piv}(d)(e)). 
As explained in the previous work \citep{zhang2019core,magnaudet2020particles}, this vortex is sourced from the baroclinic torque caused by the misalignment between the density and pressure gradients. 

A close view of the flow around the particle is presented in figure \ref{fig:piv_particle}. At the rear of the particle, an upward jet (according to a negative $u_z$) is generated along the central axis. The jet structure has been widely reported in linearly stratified fluid \citep{2000Flow,hanazaki2009jets,zhang2019core}, which was believed to be caused by the continual dragging and rupture of the lighter fluid. In our experiment, the jet is transient as the dragging and rupture of lighter fluid are temporal processes. The jet forms as the wake breaks (figure \ref{fig:piv_particle}(c)), and decays faster thereafter. As the particle bounces (figure \ref{fig:piv_particle}(g)), the jet is a dominating structure at the vicinity of the particle.
It is reasonable to deduce that the bouncing of particle is caused by the pull of the jet (as discussed in section \ref{sec:bouncing_forces}).
However, we find that the occurrence of jet is quite common for a particle settling at moderate Reynolds numbers, even without bouncing motion. The jet is actually accompanied with the aforementioned wake detachment, which is observed throughout the parameter range. As we will attempt to address that the jet is a necessary but not sufficient condition for bouncing.

\subsection{Force analysis}\label{sec:force_analysis}

As we know, the nonmonotonic motion of a particle settling in a stratified fluid is caused by a so-called `stratification drag' (noted as $F_s$). Previous studies revealed that $F_s$ is mainly contributed by two mechanisms: the buoyancy of dragged upper fluid and the force caused by a specific flow structure \citep{srdic1999gravitational,zhang2019core,higginson2003drag,2009Enhanced}.
In this section, we analyse the force acting on a bouncing particle, to further understand the mechanism of the bouncing behaviour. For the convenience of force decomposition, we present the numerical results of a settling particle with the setups matching that of the experiments in section \ref{sec:flows}. 
The simulated non-dimensional parameters are $Re_u=198$, $Re_l=26$ and $Fr=2.3$. The simulated density field is presented in figure \ref{fig:sim_process}, which exhibits the same wake structure with that in the experiment. Due to the difference of the lower Reynolds number, the bouncing behaviour is more obvious in experiment, as shown in figure \ref{fig:valid_vy}. We detail in section \ref{sec:rel} that the occurrence of bouncing depends mainly on $Re_l$. Overall, the simulation results are in agreement with the experiments.

We present the force analysis by three steps. (1) Introduce how we decompose the force. (2) Describe how we calculate the force components. (3) Present the forces acting on a bouncing particle and analyse the bouncing mechanism.

\subsubsection{Force decomposition}\label{sec:force_decomposition}

To understand the underlying physics of particle bouncing behaviour, we attempt to decompose its hydrodynamic force into different components. We start from the motion equation of a particle settling from the rest in a quiescent homogeneous fluid, which can be written as
\begin{equation}
m_p\frac{dU}{dt}=G+F_b+F_d+F_a+F_h.
  \label{eqn:homo_motion}
\end{equation}
The left-hand side is the total inertia force acting upon the particle, where $m_p$ is the particle mass. It arises due to the imbalance of forces. At the right-hand side, $G$ and $F_b$ are respectively the gravity and buoyancy forces of the particle, $F_d$ is the `steady-state' drag force at the considered time instant, $F_a$ is the inertia force of added mass, and $F_h$ is the history (Basset) force. Here, $F_d$ can be evaluated according to (\ref{eqn:Fd}) and (\ref{eqn:Cd_white}).
In the limit of potential flow, $F_a$ can be calculated as
\begin{equation}
F_a=-\frac{1}{2}  \rho_f V_p \frac{d U}{d t}.
\label{eqn:Fa}
\end{equation}
At the Stokes regime, $F_h$ has an analytic solution
\begin{equation}
F_h=-\frac{3}{2} D^2 \sqrt{\pi \rho_f \mu} \int _{-\infty} ^t \frac{\dot{U}(\tau)}{\sqrt{t-\tau}} d \tau .
\label{eqn:Fh}
\end{equation}
For the particle settling in a stratified fluid, we follow the same way of drag force decomposition as in (\ref{eqn:homo_motion}), while introducing an extra term $F_s$, accounting for the stratification effects. The motion equation becomes
\begin{equation}
m_p\frac{dU}{dt}=G+F_b+F_d+F_a+F_h+F_s.
  \label{eqn:fs}
\end{equation}
It is reasonable to further decompose $F_s$ into two components as 
\begin{equation}
F_s=F_{sb}+F_{sj},
  \label{eqn:Fs_decomposition}
\end{equation}
where $F_{sb}$ is the enhanced buoyancy caused by dragging the upper fluid to the lower layer, therefore modifying the density distributions, and $F_{sj}$ is the force caused by the induced flow structure due to the stratification, represented typically by the upward jet flow at the rear of the particle. As we have discussed in figure \ref{fig:piv_particle}, the observed jet is conjectured to be the dominant flow structure as the particle bounces. The equation of motion for a particle settling in a stratified fluid can therefore be written as
\begin{equation}
m_p\frac{dU}{dt}=G+F_b+F_d+F_a+F_h+F_{sb}+F_{sj}.
  \label{eqn:stra_motion}
\end{equation}

\subsubsection{Force calculation}\label{sec:force_estimation}
It is possible to evaluate numerically the different force components. The most convenient way is to reorganise (\ref{eqn:stra_motion}) as
\begin{equation}
m_p\frac{dU}{dt}=G+\underbrace{(F_b+F_{sb})}_{F_{static}}+\underbrace{(F_d+F_a+F_h+F_{sj})}_{F_{dynamic}},
  \label{eqn:motion_fhydro}
\end{equation}
where $F_{static}$ is the hydrostatic force caused by density stratification and its nonuniform distributions, and $F_{dynamic}$ is the force caused by the non-zero velocity field at uniform density distribution. Excluding the steady-state drag $F_d$, the sum of ($F_a+F_h+F_{sj}$) represents the force caused by unsteady flow, i.e., the flow caused by the stratification effect.
For a given density field, the hydrostatic pressure $p_s$ can be obtained by solving
\begin{equation}
\nabla p_s = \rho \bm{g}.
  \label{eqn:ps}
\end{equation}
$F_{static}$ is the integration of $p_s$ over the particle surface:
\begin{equation}
 F_{static}=-\int _S p_s \bm{n} d S ,
  \label{eqn:fstatic}
\end{equation}
where $\bm{n}$ is the unit normal at the surface. Since $F_b$ can be calculated by considering an undisturbed density profile, we have
\begin{equation}
F_{sb} = F_{static}-F_b.
  \label{eqn:fsb}
\end{equation}
The total hydro-force $F_{hydro}$ acting on the particle, i.e., $F_{static}+F_{dynamic}$, can be calculated by solving the equations (\ref{eqn:ns1}), (\ref{eqn:ns3}) and (\ref{eqn:ns4}). Then, we have
\begin{equation}
F_{dynamic} = F_{hydro}-F_{static}.
  \label{eqn:fdynamic}
\end{equation}
If we evaluate the steady-state drag $F_d$ using (\ref{eqn:Fd}) and (\ref{eqn:Cd_white}), we can write the drag component contributed by unsteady flow as
\begin{equation}
F_{sj} +F_a +F_h=F_{dynamic}-F_d.
  \label{eqn:fsj}
\end{equation}

\begin{figure}
  \centerline{\includegraphics[width=1\textwidth]{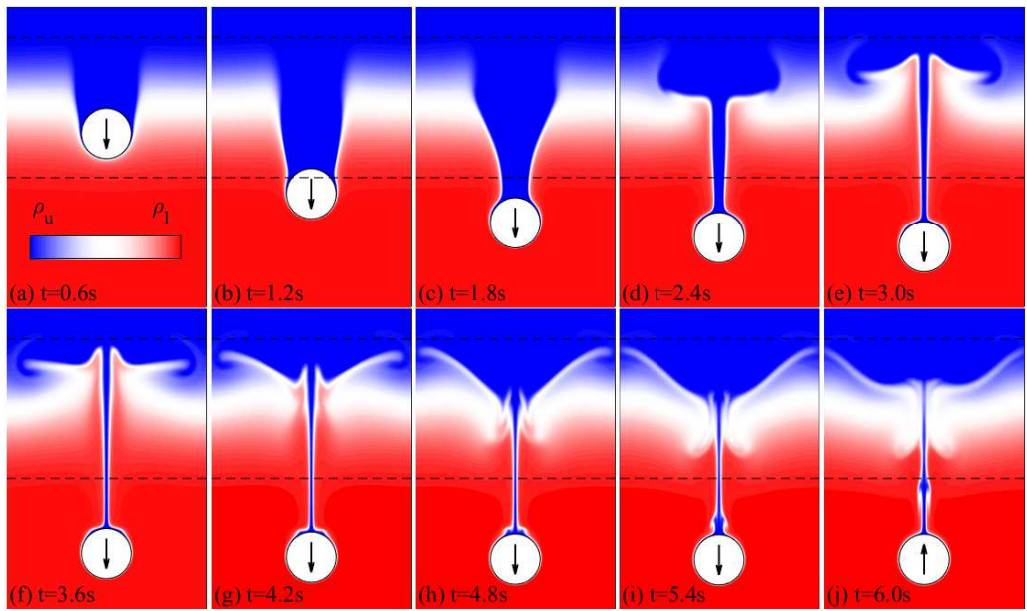}}
  \caption{Development of simulated density field of a particle settling through a density transition layer at $Re_u=198$, $Re_l=26$ and $Fr=2.3$, presented as a comparison with the experiment shown in figure \ref{fig:piv}.}
\label{fig:sim_process}
\end{figure}

\begin{figure}
  \centerline{\includegraphics[width=0.6\textwidth]{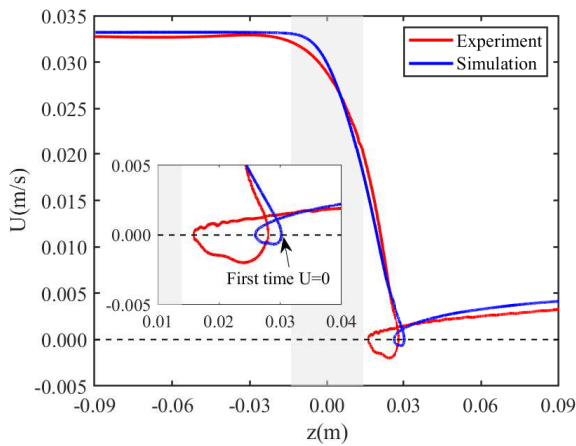}}
    \caption{Comparison between the simulated and experimental velocity profiles at $Re_u=198$ and $Fr=2.3$. Note that the lower Reynolds number for simulation is $Re_l=26$, while that for the experiment is $Re_l=20$.}
\label{fig:valid_vy}
\end{figure}

\begin{figure}
  \centerline{\includegraphics[width=1\textwidth]{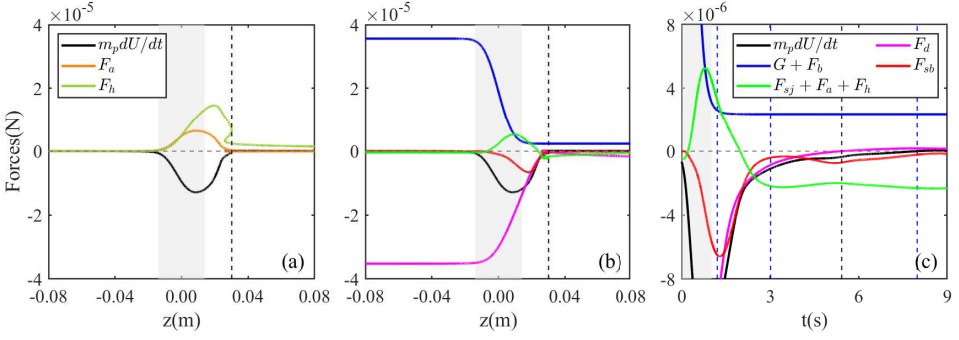}}
  \caption{Decomposed forces of a bouncing particle at $Re_u=198$, $Re_l=26$ and $Fr=2.3$, corresponding to figure \ref{fig:valid_vy}.  
  The shadow area represents the interface region. The black dashed lines show the inception of the bouncing behaviour. In figure (c), the four stages are separated by blue dashed lines, which are: wake attachment, from t=0 to the first blue dashed line; wake detachment, from the first to the second blue dashed line; transient bouncing, from the second to the third blue dashed line; finial sedimentation, from the third blue dashed line to the end.}
\label{fig:forces}
\end{figure}

\subsubsection{Forces of a bouncing particle}\label{sec:bouncing_forces}
For a particle settling through a density interface, if $U_{min}\leq 0$ (keeping in mind that the positive direction of $z-$axis points downwards), it bounces up. At the instant when the first time the particle reaches $U=0$ (see the arrow in figure \ref{fig:valid_vy}), the particle satisfies
\begin{equation}
\frac{dU}{dt}<0.
  \label{eqn:cri}
\end{equation}
Thus at $U=0$ we have
\begin{equation}
m_p\frac{dU}{dt}= \underbrace{G+F_b}_{+}+\underbrace{F_d}_{0}+\underbrace{F_a}_{+}+\underbrace{F_h}_{+}+\underbrace{F_{sb}}_{-}+\underbrace{F_{sj}}_{-}<0.
  \label{eqn:force_dirction}
\end{equation}
In figure \ref{fig:forces}(a), we plot $F_a$ and $F_h$ predicted by (\ref{eqn:Fa}) and (\ref{eqn:Fh}). We note that the solutions given by (\ref{eqn:Fa}) and (\ref{eqn:Fh}) can actually not be accurately applied to our Reynolds number ($Re_u\approx 200$). However, at least we know that $F_a$ and $F_h$ are opposite to the acceleration of the particle, i.e, positive at this time instant. 
Apparently, the necessary condition for (\ref{eqn:force_dirction}) to hold is
\begin{equation}
|F_{sb}|+|F_{sj}|>|G+F_b|.
  \label{eqn:bouncing}
\end{equation}
Whether the jet is necessary for the occurrence of bouncing depends on the magnitude of $F_{sb}$. When $|F_{sb}|\leq |G+F_b|$, the contribution of jet is necessary for the bouncing behaviour. Observed from our experiments, as discussed in section \ref{sec:bounce_process}, the particle reverses its motion direction after the wake detaches, so that $F_{sb}$ is relatively small, therefore $F_{sj}$ is determinant for the bouncing behaviour.

The decomposed force components as the functions of vertical position is plotted in figure \ref{fig:forces}(b), where ($G+F_b$) is the reduced gravity. Since $F_a$ and $F_h$ cannot be accurately calculated, we plot  ($F_{sj}+F_a+F_h$), which is the force due to the unsteady flow, and also gives the upper limit of $F_{sj}$. Note that ($F_a+F_h$) is always positive during the settling process. Apparently, before entering the transition layer, the balance between the reduced gravity ($G + F_b$) and the drag force $F_d$ makes the particle reach a constant settling velocity. As the particle enters the transition layer, the force components $m_pdU/dt$, $F_{sb}$ and $F_{sj}+F_a+F_h$ arise from their nearly zero values. We also present the time histories of different force components in figure \ref{fig:forces}(c), with the aforementioned four stages (section \ref{sec:bounce_process}) separated by three blue dashed lines. The wake buoyancy force $F_{sb}$ correlates to the volume of attached upper light fluid. It reaches a maximum at the end of the first stage, after the dragging of the wake (see figure \ref{fig:sim_process}(b) for $t=1.2$s), and before the wake detaches from the particle. The sharp rise of ($F_{sj}+F_a+F_h$) (the green line) is dominated by $(F_a+F_h)$ at the first two stages due to the sudden deceleration of the particle. 
As approaching the third stage, $F_{sb}$ becomes small since only a thin layer of light fluid left in the attached wake (see figure \ref{fig:sim_process}(e) for $t=3.0$s). In the mean time, ($F_{sj}+F_a+F_h$) appears to be the dominant drag force (to balance the reduced gravity ($G+F_b$)), arresting further settling of the particle. At the instant of bouncing occurrence, the settling velocity $U$ approaches zero (marked by the black dashed line in figure \ref{fig:forces} and see also figure \ref{fig:sim_process} between $t=5.4$s and $t=6.0$s).  At this time, $|F_{sb}| \ll |G+F_b|$, and ($F_{sj}+F_a+F_h$) plays a dominant role in balancing the reduced gravity. 
Excluding ($F_a+F_h$), the contribution from $F_{sj}$ takes the primary part of the drag force. Therefore we conjecture that the jet flow is a necessary condition for bouncing the particle up. In conclude, we prefer to the interpretation that $F_{sb}$ is the primary factor for decelerating the particle as it enters the transition layer, while $F_{sj}$ induced by the jet flow further decelerates the particle till reverses its motion as it leaves the transition layer.

\subsection{Influence of different parameters}

\begin{table}
  \begin{center}
\def~{\hphantom{0}}
\begin{tabular}{cccccccccc}
Experiment                & Test  & $\rho_u$     & $\rho_l$     & $L$      & \multicolumn{5}{c}{Minimal velocity (cm/s)}        \\
series                    & number & (kg/m$^3$) & (kg/m$^3$) & (cm) & P1     & P2     & P3    & P4    & P5    \\ \hline
\multirow{5}{*}{series 1} & 1      & 1116.52    & 1123.66    & 3.14   & -0.492 & -0.228 & 0.304 & 0.578 & 0.889 \\
                          & 2      & 1116.52    & 1123.32    & 3.01   & -0.140 & 0.237  & 0.560 & 0.780 & 1.048 \\
                          & 3      & 1116.52    & 1123.01    & 2.92   & 0.250  & 0.468  & 0.735 & 0.952 & 1.218 \\
                          & 4      & 1116.55    & 1122.71    & 2.86   & 0.415  & 0.610  & 0.858 & 1.059 & 1.302 \\
                          & 5      & 1116.52    & 1122.39    & 2.52   & 0.683  & 0.835  & 1.067 & 1.249 & 1.467 \\ \hline
\multirow{5}{*}{series 2} & 6      & 1115.05    & 1121.87    & 2.71   & -0.397 & 0.015  & 0.454 & 0.712 & 0.994 \\
                          & 7      & 1115.90     & 1121.85    & 3.22   & -0.324 & 0.114  & 0.497 & 0.739 & 1.025 \\
                          & 8      & 1116.86    & 1121.85    & 3.23   & -0.414 & -0.100 & 0.376 & 0.648 & 0.939 \\
                          & 9      & 1117.97    & 1121.95    & 3.02   & -0.337 & 0.064  & 0.475 & 0.742 & 1.022 \\
                          & 10     & 1118.63    & 1121.45    & 3.67   & -0.162 & 0.137  & 0.516 & 0.781 & 0.985 \\ \hline
\multirow{5}{*}{series 3} & 11     & 1113.93    & 1122.24    & 3.15   & -0.559 & -0.214 & 0.336 & 0.616 & 0.879 \\
                          & 12     & 1113.92    & 1122.24    & 6.72   & -0.248 & -0.087 & 0.376 & 0.651 & 0.922 \\
                          & 13     & 1113.92    & 1122.24    & 9.83   & -0.117 & 0.014  & 0.371 & 0.639 & 0.928 \\
                          & 14     & 1113.92    & 1122.24    & 11.53  & -0.035 & 0.103  & 0.435 & 0.705 & 0.954 \\
                          & 15     & 1113.92    & 1122.24    & 13.06  & 0.013  & 0.118  & 0.461 & 0.724 & 0.993 \\ \hline
\end{tabular}
  \caption{Experimental parameters and the minimal velocities.}
  \label{tab:exp_para}
 \end{center}
\end{table}

\begin{table}
  \begin{center}
\def~{\hphantom{0}}
\begin{tabular}{ccc}
Particle number  & Diameter (mm) & Density (kg/m$^3$)    \\
P1       & $10.075$        & $1121.58 - 1123.74$  \\
P2       & $10.128$        & $1121.86 - 1124.00$     \\
P3       & $10.083$        & $1122.28 - 1124.60 $  \\
P4       & $10.075$       & $1122.73 - 1124.79$  \\
P5       & $10.055$        & $1123.18 - 1125.26 $
\end{tabular}
  \caption{Particle properties.}
  \label{tab:particle}
  \end{center}
\end{table}

We carry out a parametric study by varying the controlling parameters. Three series of experiments are conducted. Each series includes five tests, as listed in table \ref{tab:exp_para}. We vary the lower fluid density $\rho_l$, upper fluid density $\rho_u$, and the interface (transition layer) thickness $L$ to get different lower layer Reynolds number $Re_l$, upper layer Reynolds number $Re_u$, and Froude number $Fr$.
For each test, five particles with slightly different properties (as listed in table \ref{tab:particle}) are released. The minimal velocity $U_{min}$ that a particle reaches for each release has also been included in table \ref{tab:exp_para}. Note that the negative $U_{min}$ implies the occurrence of bouncing.
Since the particle density varies with ambient temperature, before each test, the particle density is precisely measured to an accuracy of 0.01 kg/m$^3$.

\begin{figure}
  \centerline{\includegraphics[width=1\textwidth]{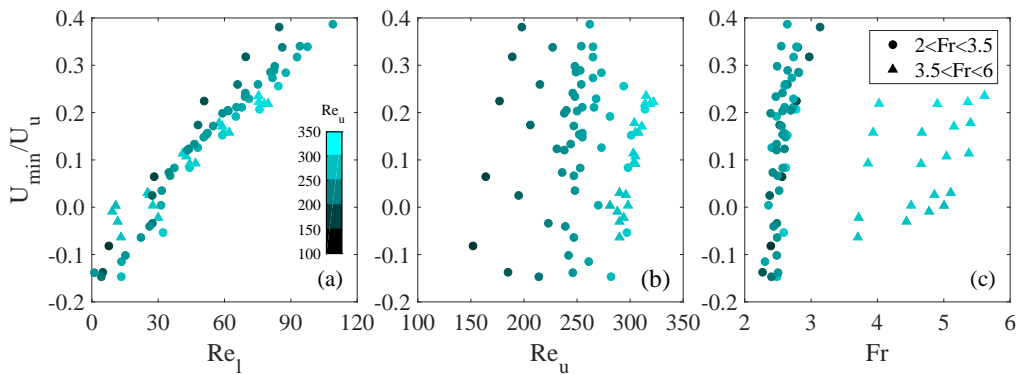}}
  \caption{The minimal velocity of all the particles in experiments versus upper, lower Reynolds numbers and Froude number. $U_{min}/U_u<0$ represents a bouncing behaviour. }
\label{fig:umin_reu_rel_fr}
\end{figure}

The non-dimensional minimal velocities versus  $Re_l$, $Re_u$ and $Fr$ for all the collected experimental data are shown in figure \ref{fig:umin_reu_rel_fr}. 
Clearly, the minimal velocity ($U_{min}/U_u$) correlates more strongly to the lower Reynolds number ($Re_l=1 \sim 109$) than the upper Reynolds number ($Re_u=152 \sim322$) and the Froude number ($Fr$=$2.3 \sim 5.6$). The bounce is seen to occur as $Re_l$ is smaller than about 30. 

\subsubsection{Lower Reynolds number $Re_l$}\label{sec:rel}
\begin{figure}
  \centerline{\includegraphics[width=1\textwidth]{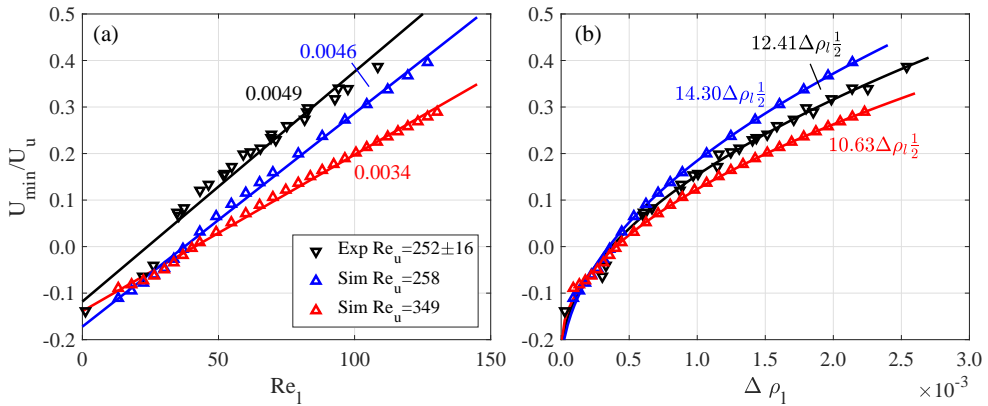}}
  \caption{The variations of non-dimensional minimum velocity over (a) $Re_l$ and (b) $\Delta \rho_l$. The Froude number are $Fr=2.6\pm0.2$ for experiment and $Fr=2.6$ for simulation.}
\label{fig:umin_Rel}
\end{figure}

To further investigate the relationship between the minimal velocity and the lower Reynolds number, the experimental data from series 1, where the upper Reynolds number and Froude number are $Re_u=252\pm16$ and $Fr=2.6\pm0.2$ respectively, are plotted over the lower Reynolds number $Re_l$ in figure \ref{fig:umin_Rel}(a). The numerical results at $Re_u=258$ and $Re_u=349$ with the fixed Froude number $Fr=2.6$ are also plotted for comparison.
From the trend, the experimental and numerical results are consistent, both indicating that $U_{min}/U_u $ increases linearly with $Re_l$. We note that for $Re_u \sim252$, the values of $U_{min}/U_u$ in the experiments are uniformly higher than that of the simulations, which is possibly caused by the incomplete development of $U_l$ in the experiments. After passing through the interface, a long distance is required for the particle to reach a steady state velocity. Such that the velocity measured at the instant when the particle leaves the viewing window in our experiments might be smaller than the fully developed $U_l$ in the simulations, resulting a smaller $Re_l$.
We can fit the data shown in figure \ref{fig:umin_Rel}(a) by the lines expresses as \begin{equation}
\frac{U_{min}}{U_u}=c_1Re_{l}+c_2.
  \label{eqn:umin_rel}
\end{equation}
The linear regression gives $c_1=0.0049$ and $c_2=-0.1181$ for the experiments ($Re_u=252\pm16$), and $c_1=0.0046$ and $c_2=-0.1720$ for the numerical results ($Re_u=258$). Critical lower Reynolds numbers $Re^ \ast_{l}=23.9$ and $37.5$ are identified respectively for these two lines, i.e., the position intersected with the horizontal axis, or when $U_{min}=0$. It is easy to understand that the undetermined parameters $c_1$ and $c_2$ are dependent on $Re_u$ and $Fr$. Evidenced from \ref{fig:umin_Rel}(a), with a higher upper  Reynolds number $Re_u=349$, the fitting line gives a smaller slope compared with both the experiments and simulations with the smaller Reynolds numbers. However, it is interesting to find that the critical lower Reynolds numbers $Re^\ast _{l}$ obtained from the numerical simulations with different $Re_u$ are very close.

In our experiments, since the terminal Reynolds numbers are not known $a$ $priori$, we practically adjust the fluid density to get varied Reynolds numbers. Previous works show that the bouncing behaviour is related to the fluid density and density ratio \citep{camassa2022critical,doostmohammadi2014reorientation}. 
Thus the relationship between the minimal velocity and the density ratio should also be investigated. Here, we define the density ratio as $\Delta \rho_l=(\rho_p-\rho_l)/\rho_l$.
Since $Re_l$ is correlated to the density difference, the relationship between $U_{min}$ and $\Delta \rho_l$ can be derived from (\ref{eqn:umin_rel}).
For a particle settling to a steady-state velocity in a homogeneous fluid, the drag force balances the reduced gravity, satisfying the following expression:
\begin{equation}
F_d=G-F_b. 
  \label{eqn:fd_balance}
\end{equation}
Using the definition of $F_d$ in equation (\ref{eqn:Fd}) and substituting $U_l= \nu_l Re_l  /D$ into (\ref{eqn:fd_balance}), we get the following expression between density ratio and Reynolds number: 
\begin{equation}
\Delta \rho_l=\frac{3C_{dl} \nu_l ^2 Re_{l}^2 }{4 g D^3},
  \label{eqn:rho_re}
\end{equation}
where $C_{dl}$ is the steady-state drag coefficient in the lower layer.
Equations (\ref{eqn:umin_rel}) and (\ref{eqn:rho_re}) yield
\begin{equation}
\frac{U_{min}}{U_u}=c_1 \sqrt{\frac{4g D^3}{3 \nu_l ^2 C_{dl}} } \Delta \rho_l ^{\frac{1}{2}}+c_2.
  \label{eqn:umin_deltarho}
\end{equation}
We find that the power law fitting in the form 
\begin{equation}
\frac{U_{min}}{U_u}=c_3 \Delta \rho_l ^{\frac{1}{2}}+c_4,
  \label{eqn:umin_deltarho_fit}
\end{equation}
is appropriate to describe the dependence of $U_{min}/U_u$ on $\Delta \rho_l$, as shown in figure \ref{fig:umin_Rel}(b) (recompiling the data from figure \ref{fig:umin_Rel}(a)).

\begin{figure}
  \centerline{\includegraphics[width=1\textwidth]{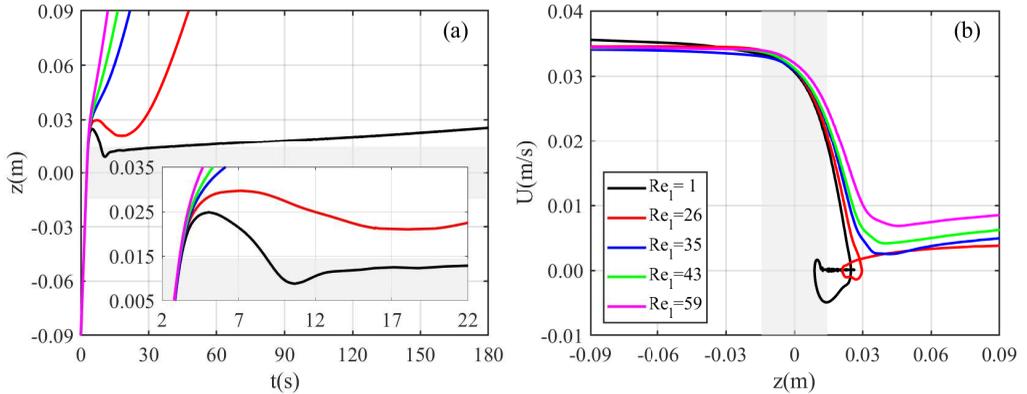}}
  \caption{(a) Time trajectories of particles at five different lower Reynolds numbers. The nonmonotonic trend indicates a bouncing behaviour. (b) The velocity profiles of particles at five different lower Reynolds numbers. The grey region refers to the density transition layer. $Fr$ and $Re_u$ vary slightly within $Fr=2.4 \sim 2.5$ and $Re_u=236 \sim 246$.}
\label{fig:yv_Rel}
\end{figure}

The trajectories and velocity profiles of particle $P1$ from the experiment series 1 are plotted in figure \ref{fig:yv_Rel}. It is clearly shown in figure \ref{fig:yv_Rel}(a) that the sedimentation time is dramatically prolonged by the bouncing behaviour. Although the particle enters the interface with the same velocity, the slight variation of $Re_l$ significantly alters the behaviour.
For a relatively large lower layer Reynolds number, e.g. $Re_l=59$, though a minimum velocity is observed in figure \ref{fig:yv_Rel}(b), the particle keeps descending unidirectionally after it passes through the transition layer. We emphasise that all particles have larger density than the fluid in the tank at all attitudes. As $Re_l$ decreases, crossing a critical value, i.e. $Re^ \ast_{l}$, as we have discussed above, the particle reverses its direction of motion and ascends for a transient time scale (see figure \ref{fig:yv_Rel}(a) for $Re_l=26$ and $Re_l=1$). This bouncing phenomenon is represented by a much deeper and negative minimum velocity shown in the depth vs velocity plot of figure \ref{fig:yv_Rel}(b). Here, $Re_l=1$ refers to a very extreme case, when the particle density ($\rho_s=1123.69$ kg/m$^3$) is near the density of the bottom layer ($\rho_l=1123.66$ kg/m$^3$). In this case, the particle experiences an extraordinarily long transient time scale to reach the terminal velocity of the bottom layer. An explanation for this long transient, given by the previous study \citep{abaid2004internal}, is that once the plume of upper layer fluid has shed, there still exists around the particle a small boundary layer of upper fluid which diffuses exceptionally slowly due to the very long diffusion of time of salt in water and the absence of a strong turbulence diffusion in this low-speed flow state. This is evidenced by our experimental results in figure \ref{fig:bounce}(p$\sim$t), which show clearly that a few light fluid remains at the particle surface after the bouncing.

It is worthy mentioning some previous studies, where the bouncing behaviour was not observed. \citet{srdic1999gravitational} presented the time trajectories of particles obtained by a series of experiments (see their figure 9). They reported the noticeable decrease of velocity within the transition layer, however without finding reverse motion of the particle. It can be possibly explained from two aspects. First, in their experiments, the upper fluid was a mixture of ethyl alcohol and water, which diffuses faster than salty water. Thus the lighter upper fluid dragged by the particle could adjust to the surrounding fluid immediately and weaken the deceleration of the particle. Second, and more importantly, their examined upper Reynolds numbers fall in the range $0.7\leq Re_u\leq23$, much lower than the current studies. As we will discuss later, the upper Reynolds number is also one of the influencing factors for the occurrence of bouncing behaviour. Actually, the second explanation can also be related to a deeper physical mechanism, the rear buoyant jet, which disappears when $Re_u$ is low. As evidenced from the work by \citet{2009Enhanced}, there was no sign of such a jet as $Re_u\sim1$. 

\subsubsection{Froude number $Fr$}\label{sec:fr}
We now address the effects of Froude number. Referring to the experiment series 3 presented in table \ref{tab:para}, we fix both the upper and lower layer fluid densities, varying only the transition layer thickness. This is achieved by releasing the particles in the same tank of fluid with a time interval of 24 hours. The diffusion of salt along such a long time scale gives the variations of transition layer thickness in the range $L=$3.15 cm - 13.06 cm, resulting in various Froude numbers. 

\begin{figure}
  \centerline{\includegraphics[width=1\textwidth]{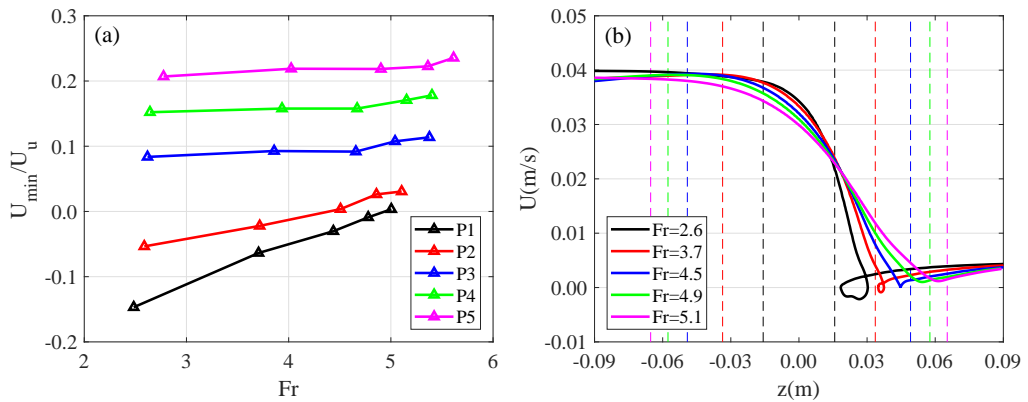}}
  \caption{(a) The non-dimensional minimum velocity versus Froude number for five particles. The lower and upper Reynolds numbers are: P1, $Re_l=12,Re_u=286$; P2, $Re_l=28,Re_u=295$; P3, $Re_l=44,Re_u=303$; P4, $Re_l=59,Re_u=306$; P5, $Re_l=77,Re_u=317$. (b) The velocity profiles of particle P2 versus the vertical position at five Froude numbers. The Vertical dashed lines indicate the bounds of the density transition layers.}
\label{fig:umin_fr_yv_p2}
\end{figure}

We plot the non-dimensional minimum velocities for different particles over the Froude number in figure \ref{fig:umin_fr_yv_p2}(a). Note that the particle density increases from P1 to P5, therefore both the upper and lower Reynolds numbers increase from P1 to P5. The general trend is a nearly monotonous increase in the non-dimensional minimal velocity along with the increasing Froude number, i.e., as the transition layer becomes thicker. Comparing between the tests of different particles, we can see an overall increase in the minimum velocity when the particle density increases, or equivalently when the Reynolds number increases. Another trend is the less role of Froude number playing in the increase of minimum velocity when the particle density is large (see P5 in figure \ref{fig:umin_fr_yv_p2}(a)). In contrast, for the lighter particles, such as P1 and P2, the influence of Froude number is remarkable. For example, the behaviour of particle P2 is altered from bouncing to unidirectional settling as the Froude number increases from $Fr=2.6$ to $5.1$. Note that in the tests of P2, the lower Reynolds number ($Re_l=28) $ is very close to the critical Reynolds number $Re_l ^ \ast$. In figure \ref{fig:umin_fr_yv_p2}, for P3, P4 and P5, no bouncing phenomenon is recorded.

Specifically, we present the velocity profiles of particle P2 at different Froude numbers in figure \ref{fig:umin_fr_yv_p2}(b). In these tests, the Froude number significantly alters the evolving profile of particle settling velocity. We emphasise that the bouncing motion occurs after the particle leaves the transition layer (see $Fr$=2.6 and 3.7 in figure \ref{fig:umin_fr_yv_p2}(b)), while the minimum velocity is reached within the transition layer for those tests without bouncing (see $Fr$=4.5, 4.9 and 5.1 in figure \ref{fig:umin_fr_yv_p2}(b)). There is an interesting phenomenon we want to report. In figure \ref{fig:umin_fr_yv_p2}(b) we observe that the particle restores to a higher settling velocity for the lower Froude number tests than that with high Froude numbers. It might be caused by the more pronounced diffusion induced by the jet flow. Clearly, the jet flow can enhance the turbulence diffusion and diffuse the remaining upper layer fluid attached on the particle to its ambient lower layer fluid.

\subsubsection{Upper Reynolds number $Re_u$}\label{sec:reu}
\begin{figure}
  \centerline{\includegraphics[width=1\textwidth]{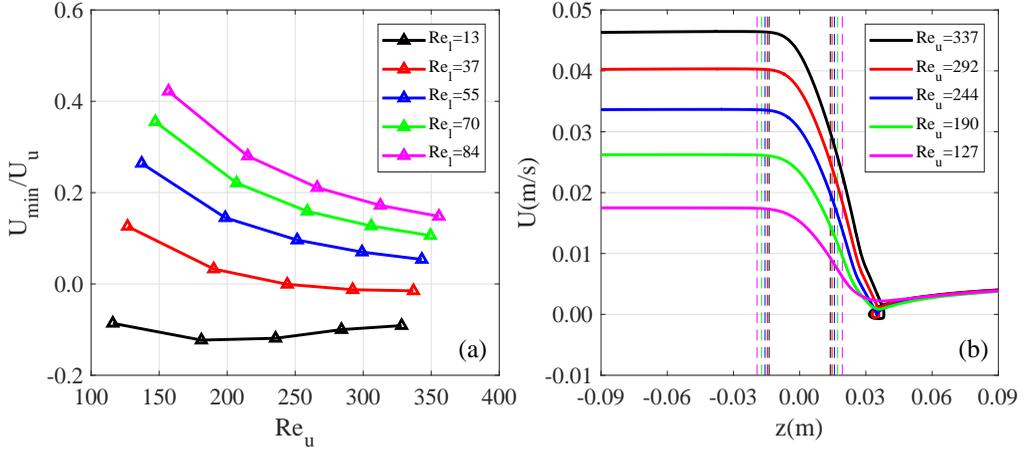}}
  \caption{(a) The non-dimensional minimal velocity versus the upper Reynolds number at five lower Reynolds numbers, with $Fr=2.6$. (b) The velocity verses vertical position corresponding to $Re_l=37$ in (a).}
\label{fig:yv_Reu}
\end{figure}

We next investigate the effects of upper Reynolds number $Re_u$ by numerical simulations.
In figure \ref{fig:yv_Reu}(a), the non-dimensional minimal velocity versus $Re_u$ at five different values of $Re_l$ are presented, at a fixed Froude number $Fr=2.6$. The trend is clear: the minimal velocity increases with the increased lower Reynolds number $Re_l$, while at a fixed $Re_l$ the minimal velocity decreases with the increase of $Re_u$ except for $Re_l=13$, where all minimal velocities are negative, indicating the occurrence of bouncing for all cases. We examine a special case, $Re_l=37$, when the minimum velocity becomes negative as the upper Reynolds number rises to $Re_u=292$. In figure \ref{fig:yv_Reu}(b), we plot the variations of settling velocity with depth for this special case. Despite the distinct differences in entering velocities of the particle among different tests, their minimal velocities are considerably resembling. Also, they reach minimal velocities at nearly the same depth. Note that the thickness of transition layer varies little among different tests. From figure \ref{fig:umin_reu_rel_fr} we understand that $Re_l=37$ is near the critical Reynolds number for the occurrence of bouncing, therefore this special case is very sensitive to the parameters, such as the upper Reynolds number $Re_u$. As demonstrated in figure \ref{fig:yv_Reu}(b), the particle moves unidirectionally for $Re_u=127, 190$ and 244, while reverses its motion direction for $Re_u=292$ and 337.

We thus stress that the upper Reynolds number $Re_u$ plays a non-negligible role only when the bottom layer Reynolds number $Re_l$ is close to its critical value for bouncing. However, we note that the currently studied upper Reynolds numbers, in both experiments and numerical simulations, are kept at a certain order of $Re_u \sim O(100)$, while a much lower $Re_u$, e.g. in the order $\sim O(1)$ will give no sign of bouncing, as we have discussed in section \ref{sec:rel}.

\subsubsection{Identification of bouncing regime in ($Re_u$, $Fr$) space}\label{sec:reu_fr}

\begin{figure}
  \centerline{\includegraphics[width=1\textwidth]{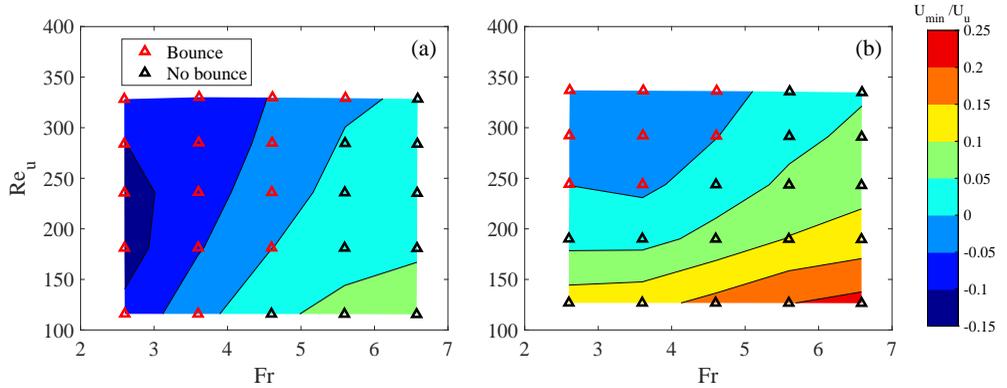}}
  \caption{Maps of non-dimensional minimal velocities of the settling particle at (a) $Re_l=13$, and (b) $Re_l=37$.}
\label{fig:umin_cotour}
\end{figure}

Since the bouncing behaviour is primarily determined by the lower Reynolds number $Re_l$, while the other two parameters, $Re_u$ and $Fr$ can also play their roles, as we have discussed in section \ref{sec:fr} and \ref{sec:reu}, it is possible to give a better visualisation by examining their combined effects. For example, in figure \ref{fig:umin_cotour}, we draw two maps in the parametric space of $Re_u$ and $Fr$ for two selected lower Reynolds numbers $Re_l=13$ and $Re_l=37$, with the contours representing their minimal velocities. Clearly, the bouncing phenomenon occurs at the higher $Re_u$ and lower $Fr$, i.e. the upper-left regimes of the maps.  
The direct comparison between figure \ref{fig:umin_cotour}(a) and (b) indicates that the bouncing regime shrinks for the higher value of $Re_l$, demonstrating again that $Re_l$ is the dominant parameter. When $Re_l$ is low, the particles are more prone to bounce after passing through the transition layer.

\begin{table}
  \begin{center}
\def~{\hphantom{0}}
\begin{tabular}{cccccc}
~     & $Fr=2.6$ & $Fr=3.6$ & $Fr=4.6$ & $Fr=5.6$ & $Fr=6.6$  \\
$Re_u=349$ & 41.0    & 46.2   & 44.5   & 25.5   & $-$      \\
$Re_u=305$ & 39.6    & 43.7   & 39.4   & 14.4    & $-$      \\
$Re_u=258$ & 37.5   & 40.0   & 31.9   & $-$      & $-$       \\
$Re_u=207$ & 34.6   & 34.2   & 21.8  & $-$      & $-$       \\
$Re_u=147$ & 28.9   & 24.3   & $-$      & $-$      & $-$      
\end{tabular}
  \caption{Critical lower Reynolds number $Re_{l}^\ast$ in the ($Re_u$, $Fr$) space.}
  \label{tab:relh}
 \end{center}
\end{table}

Simulating in the same ($Re_u$, $Fr$) parametric space for different values of $Re_l$, with a lower limit of $Re_l=13$, we summarise the critical lower Reynolds numbers $Re^ \ast _{l}$ in table \ref{tab:relh}. We note that those with $Re_l<13$ are not presented in this table. We see that $Re^ \ast _{l}$ varies from $14.4$ to $46.2$ depending on different combinations of $Re_u$ and $Fr$.
We should point out that the variations of $Re^ \ast _{l}$ with $Fr$ at a fixed $Re_u$ are not always monotonous, particularly when $Re_u$ is high.

\subsection{Discussion on the strength of buoyant jet}\label{sec:jet}

\begin{figure}
  \centerline{\includegraphics[width=0.55\textwidth]{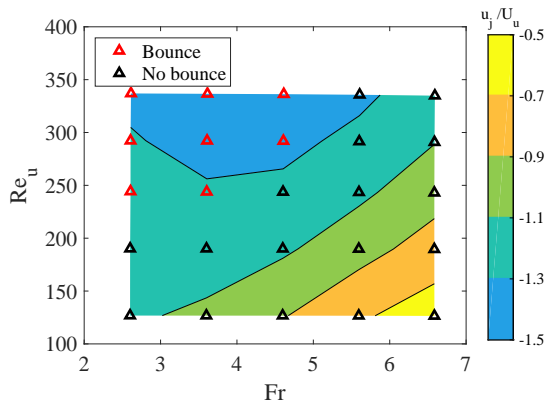}}
  \caption{Jet velocities with maximum magnitudes in $Re_u$-$Fr$ space at $Re_l=37$. The negative value refers to an upward jet.}
\label{fig:jet}
\end{figure}

It is worth examining the jet flow by quantifying its strength. Corresponding to figure \ref{fig:umin_cotour}(b), we present the maximum jet velocity for each combination of $Re_u$ and $Fr$ in figure \ref{fig:jet}. Here, the lower Reynolds number is fixed at $Re_l=37$. The maximum magnitude of jet velocity is identified at each test, which is scaled by the upper layer terminal velocity $U_u$. This non-dimensional jet velocity $u_j/U_u$ increases in its magnitude with the increased $Re_u$ while with the decreased $Fr$, indicating that the jet strength becomes stronger as the inertial effect becomes dominant, as well as the stratification becomes stronger. Not surprisingly, we find that the strong jet region exhibited in figure \ref{fig:jet}, the upper-left corner, matches well with the bouncing regime shown in figure \ref{fig:umin_cotour}(b). In other words, the upward jet flows correlate to the occurrence of bouncing behaviour. 

\section{Conclusions}\label{sec:conclusions}

In the present study, we carry out comprehensive experimental tests and numerical simulations on a spherical particle settling through a density stratified fluid, focusing on revealing the physical mechanisms for bouncing behaviour, or reverse motion, as the particle passes through the transition layer. Our study covers a wide range of parameters, with the lower layer Reynolds number $1 \leq Re_l\leq 125$, the upper layer Reynolds number $115 \leq Re_u\leq 356$, the Froude number $2 \leq Fr\leq 7$, and the Prandtl number $Pr \approx 700$ for a salinity-stratified fluid.

First, we decompose the forces acted on the particle into different components, and correlate them to the flow structure. We find that the particle experiences four sequential stages as it settles, in the order of wake attachment, wake detachment, transient bouncing and final sedimentation. Two mechanisms are identified, which contribute to the drag enhancement. First, the buoyancy of attached upper, lighter, fluid and the second, the force caused by a specific flow structure, the buoyancy jet flow.
At the first two stages, the force component $F_{sb}$ due to the attached upper fluid in the wake contributes mostly to the drag enhancement. While, at the third stage, most of the upper fluid has detached from the particle, thus $F_{sb}$ becomes less significant. Instead, the force component $F_{sj}$ induced by the jet flow, caused by the rapture of the wake, appears to be dominant. This jet flow is evidenced from our experimental measurements. We conjecture that the jet flow is a necessary condition for the occurrence of bouncing motion.

Then, we investigate respectively the influence of $Re_l$, $Re_u$ and $Fr$. We monitor the minimal settling velocity of the particle, of which a negative value indicates a bouncing motion of the particle, hence the bouncing regimes can be clearly identified in the parametric spaces. We find that the lower Reynolds number $Re_l$ is the determinant factor. In our experiments, the bouncing motion is found to occur below a critical lower Reynolds number around $Re^ \ast _{l}=30$. In the numerical simulations, the highest value for this critical number is $Re^ \ast _{l}=46.2$, limited in the currently studied parametric ranges. Moreover, by examining the jet flow by quantifying its strength, we find a consistency between the maximum magnitude of jet velocity in the flow fields and the minimal settling velocity for the particle, plotted in a same ($Re_u$, $Fr$) space, demonstrating the significance of jet flow on the particle's bouncing motion.

The study on bouncing behaviour of particles settling through a density stratified fluid is clearly still far from comprehensive. For example, it will be of interest to consider a cluster of particles, of which the interactions between particles would lead to more complicated and interesting settling behaviours. Also, particles with irregular shapes can be considered. Both can represent more closely the real situations, such as the aggregation of marine snow.

\section*{Acknowledgements}\label{sec:ack}
\begin{acknowledgments}
This research has been supported by the National Natural Science Foundation
of China (Grant No: 11922212) and the Fundamental Research Funds for the Zhejiang Provincial Universities (Grant No: 2021XZZX017). S.W. gratefully acknowledges the hospitality of the Department of Applied Mathematics and Theoretical Physics, University of Cambridge.
\end{acknowledgments}


\bibliographystyle{jfm}

\end{document}